# Title: Why is the superconducting $T_c$ so high in rare-earth-doped $CaFe_2As_2$?

**Authors:** F. Y. Wei,[1] B. Lv,[1] L. Z. Deng,[1] J. K. Meen,[2] Y. Y. Xue,[1] C. W. Chu[1,3]*

**Affiliations:**

[1]Department of Physics and Texas Center for Superconductivity, University of Houston, Houston TX 77204-5002.

[2]Department of Chemistry and Texas Center for Superconductivity, University of Houston, Houston TX 77204-5002.

[3]Lawrence Berkeley National Laboratory, Berkeley CA 94720.

*Corresponding author. E-mail: cwchu@uh.edu

**Abstract**: In rare-earth doped single crystalline $CaFe_2As_2$, the mysterious small volume fraction which superconducts up to 49 K, much higher than the bulk $T_c$ ~ 30s K, has prompted a long search for a hidden variable that could enhance the $T_c$ by more than 30% in iron-based superconductors of the same structure. Here we report a chemical, structural, and magnetic study of $CaFe_2As_2$ systematically doped with La, Ce, Pr, and Nd. Coincident with the high $T_c$ phase, we find extreme magnetic anisotropy, accompanied by an unexpected doping-independent $T_c$ and equally unexpected superparamagnetic clusters associated with As vacancies. These observations lead us to conjecture that the tantalizing $T_c$ enhancement may be associated with naturally occurring chemical interfaces and may thus provide a new paradigm in the search for superconductors with higher $T_c$.

**One Sentence Summary:** A systematic study on rare-earth doped $CaFe_2As_2$ single crystals has provided the best evidence for naturally occurring interface-enhanced $T_c$ detected to date.

**Main Text:** The discovery of superconductivity in 2008 in LaFeAs(O,F) with transition temperature ($T_c$) of 26 K has generated abundant excitement for theoretical and practical reasons (*1*). The unusual presence in this compound of Fe, which is antithetic to superconductivity, let alone high $T_c$ superconductivity, can provide a unique opportunity for examining the role of magnetism in high temperature superconductivity. Furthermore, the existence of a large number of other pnictides and chalcogenides of the same structural type may offer a new path to even higher $T_c$ superconductivity. Indeed, in the ensuing years, numerous Fe-based layered compounds have been found to display bulk superconductivity upon doping or applied pressure, with a $T_c$ as high as 57 K in SmFeAs(O,F) (*2*). However, filamentary superconductivity up to 25 K has also been detected unexpectedly in undoped single crystals of $AFe_2As_2$ (A122) at ambient pressure, where A is an alkaline earth element Ca (*3*), Sr (*4*), or Ba (*5*). Especially remarkable is the recent report of non-bulk superconductivity in Pr-doped Ca122 [(Ca,Pr)122] with an onset transition temperature $T_c$ as high as 49 K, via resistive, magnetic, and calorimetric measurements (*6-7*). This $T_c$ of 49 K is higher than that of any known compounds consisting of one or more of its constituent elements of Pr, Ca, Fe, and As in (Ca,Pr)122 at ambient or under pressures, or that of any other members of the 122-family through doping or pressure (supplementary online text S1). In addition, such a high $T_c$ exceeds that predicted for any phonon-mediated superconductors (*8*). However, the crucial question remains: why is the superconducting $T_c$ so high in rare-earth-doped Ca122? Here we will provide chemical, structural, and magnetic evidence that the elevated $T_c$ is an interface effect, and furthermore suggest that such interface superconductivity may serve as a more general route to significant $T_c$-enhancements in other Fe-based superconductors.

We have investigated the evolution of superconductivity in $CaFe_2As_2$ systematically doped with rare earth elements R = La, Ce, Pr, and Nd. Single crystal $(Ca_{1-x}R_x)Fe_2As_2$ [(Ca,R)122] was grown by the self-flux technique in an Ar-atmosphere with x up to the respective solubility limit. Wavelength-dispersive-spectrometry (WDS) analyses show all are chemically homogeneous to the scale defined by $|\Delta x| < 0.005$ across the sample surface with a spatial resolution of 1 μm. Recent STM study revealed the absence of Pr-clustering in the atomic scale in (Ca,Pr)122 (*9*). However, the observations cannot rule out possible mesoscopic-scale defect-structures between the micro- and nano-lengths. Although some (Ca,R)122 crystals undergo a collapse-phase-transition (*10*), all are superconducting (supplementary online text S2). The superconducting crystals display two sequential nonbulk superconducting transitions with different field responses and different but large magnetic anisotropies, as exemplified by $(Ca_{0.87}Pr_{0.13})122$ in Figs. 1A-D, using the Quantum Design MPMS and PPMS. For the high temperature transition, the $T_c$ varies between $T_{c1}$ = 49 K (for Pr) and 42 K (for Nd); and for the low temperature transition, $T_{c2}$ alters between 24 K (for Pr) and 14 K (for Nd) (supplementary online text S2). These high $T_{c1}$ values exceed the highest $T_c \sim$ 38 K for the bulk superconducting state induced in the 122 family by doping or pressure (*11-12*). It is evident that the appearance of these unusually high $T_c$s of the high temperature transition in (Ca,R)122 are not affected by the ~ 7%-spread of the R-ionic radii nor the presence/absence of the collapse phase, and hence cannot be attributed to the associated strain. Because of the slightly higher $T_c$ for R = Pr, only the (Ca,Pr)122 data will be presented and discussed hereafter. $(Ca_{1-x}R_x)122$ crystals for R = Sm and Gd were also prepared under the same conditions as others. As expected, they fail to enter the Ca122 crystals. These crystals together with the growth flux do not show any superconductivity above 10 K, except for a trace of superconductivity below 10 K attributable to the undoped Ca122 (supplementary online text S2). This rules out the possible minute $RFeAsO_{1-\delta}$-inclusion to be the culprit for the high $T_c$ nonbulk superconductivity detected in (Ca,R)122.

The doping effects on the two discrete $T_c$s of the two superconducting transitions and the superconducting volume fraction *f* at 5 K are shown in Figs. 2A and 2B for $(Ca_{1-x}Pr_x)122$ throughout the superconducting region $0.06 \leq x \leq 0.13$. In the present work, we have taken the 700 Hz ac magnetic susceptibility along the *ab*-plane ($4\pi\chi_{ac}^{ab}$) as *f* to reduce the uncertainty in demagnetization correction and to avoid the magnetic background interference of the samples. In addition, the low field ac magnetic susceptibility along the *ab*-plane at this frequency is found to be the same as its dc value. As Fig. 2B shows, while *f* increases continuously with x, it reaches a value of ~ 3.3% at the solubility limit. The nonbulk nature of the superconducting state is clear. At the same time, superconductivity takes place in two transitions with two discrete $T_c$s. The high temperature transition with $T_{c1} \sim$ 49-47 K appears abruptly at x ~ 0.08 and remains so up to x ~ 0.13, while the low temperature transition with $T_{c2} \sim$ 28-24 K appears suddenly at x ~ 0.06 and throughout the superconducting region up to x ~ 0.13, as shown in Fig. 2A. This suggests that the high $T_c$ is not sensitive to x and cannot result from change of carrier density due to a conventional optimal Pr-doping in small segregated regions of the crystals.

To explore the superconducting morphology of the (Ca,Pr)122 crystals, we have measured the ac magnetic susceptibility at 700 Hz of the superconducting (Ca,Pr)122 samples in different fields at 25 K and 5 K. We have chosen ac over dc magnetic susceptibility for the field-effect study because it gives a better control of the low measuring field below 1 Oe wanted, in addition to those advantages mentioned above. The ac magnetic susceptibilities along the ab-plane ($4\pi\chi_{ac}^{ab}$) and along the c-axis ($4\pi\chi_{ac}^{c}$) of nominal $(Ca_{0.87}Pr_{0.13})Fe_2As_2$ single crystals in a total field $H_t$ between $10^{-2}$ and $5\times10^3$ Oe are measured and displayed in Figs. 3A-D, where $H_t = H_{ac} + H_{dc}$ with $H_{ac}$ being the ac-exciting-field and $H_{dc}$ the dc-biased-field. They exhibit unusual step responses to fields. Fig. 3A shows $4\pi\chi_{ac}^{ab}(H_t)$ at 25 K above $T_{c2}$. As $H_t$ increases, $f^{ab}(H_t) \equiv$ the magnitude of the diamagnetic susceptibility $|4\pi\chi_{ac}^{ab}(H_t)|$ is initially constant for $H_t \leq H_{t1} \sim 5$ Oe, decreases at $H_t \geq H_{t1}$, and finally levels off at $H_{t2} \geq 500$ Oe up to 5 kOe. The much larger $f^c(H_t) \equiv |4\pi\chi_{ac}^{c}(H_t)|$ at 25 K before demagnetization correction in Fig. 3B exhibits a similar behavior as $H_t$ increases, i.e. it is also a constant at low $H_t$ and decreases above $H_{t1} \sim 0.3$ Oe continuously but without leveling off to our highest measuring field of 5 kOe. The observations also give an unexpectedly large magnetic anisotropy $\eta = \chi_{ac}^{c}/\chi_{ac}^{ab}$ whose exact value depends on the field, temperature, and the sample. For instance, at $H_t \sim 0.1$ Oe, $\eta \sim 57$ at 25 K and 42 at 5 K for this sample in contrast to the sample geometric anisotropy of $\sim 4$. Grossly similar field effects on $4\pi\chi_{ac}^{ab}(H_t)$ and $4\pi\chi_{ac}^{c}(H_t)$ at 5 K below $T_{c2}$ have also been observed, as shown in Figs. 3C-D.

The observations shown in Figs. 3A-D are reminiscent of flux penetration into a granular superconductor (*13*). Therefore, we propose in Fig. 4 the model for the schematic morphology to account for the nonbulk superconductivity in the (Ca,R)122 crystals with unusually high $T_c$ and large magnetic anisotropy. We envision the crystals to comprise sparsely populated, highly anisotropic superconducting domains that consist of Josephson Junction Arrays (JJAs) (*14*) of superconducting platelets with possible stacks of FeAs-layers of different defect densities (for reasons to be described later) and thus with different electronic properties to form interfaces responsible for the observed $T_c$ enhancement. The unusual field-dependences of $f^{ab}$ and $f^c$ at 25 K (and also at 5 K) can thus be understood readily in terms of this model as follow: when $H_t$ increases, $f^{ab}$ (Fig. 3A) is initially a constant up to $H_t = H_{t1} \sim 10$ Oe when the domains are completely shielded by the JJAs; decreases as $H_t > H_{t1}$ when Josephson vortices form and enter to the domains between the platelets; and finally stops decreasing between $H_{t2} \sim 500$ Oe up to 5 kOe when all weak-links between the superconducting platelets inside the domains are decoupled but before Abrikosov vortices form and enter the superconducting platelets due to the large effective lower-critical-field $H_{c1}$ of the platelets (> 5 kOe) associated with the large $\eta$ and the small dimension of the platelets along the *ab*-direction (*15*). Similarly, as for $f^c$ displayed in Fig. 3B, it is also constant for $H_t < H_{t1} \sim 4 \times 10^{-1}$ Oe, and starts to decrease continuously for $H_t > H_{t1} \sim 4 \times 10^{-1}$ Oe when Josephson vortices begin to form and enter the domain between the platelets up to 5 Oe. The continued decrease of $f^c$ for $H_t \geq 5$ Oe without reaching a constant as for $f^{ab}$ can be attributed to the combined formations of the Josephson vortices above $H_{t1}$ to destroy the JJAs and of the Abrikosov vortices above $H_{c1}$ to enter the superconducting platelets. The smaller $H_{t1}$ and the smaller apparent $H_{c1}$ along the c axis can be attributed to the edge effect of the highly anisotropic superconducting platelets.

The field dependence of $f^{ab}$ (Fig. 3A) allows us to estimate the thickness $d$ of the platelets. Assuming that the length of the platelet is large compared with $d$, the suppression of the superconducting screening $S = 1/\left[1 - \frac{2\lambda}{d} \cdot \tanh\left(\frac{d}{2\lambda}\right)\right] \sim f^{ab}(<H_{t1})/f^{ab}(>H_{t2}) = 6$, where $\lambda$ is the penetration depth (*16*). For known Fe-pnictide superconductors, $\lambda$ varies between 0.2 and 0.6 μm (*17*). $d$ for the superconducting platelets is thus on the order of sub-microns, consistent with the dimension-induced, unusually large $H_{c1} > 5k$ Oe along the *ab*-direction suggested in Fig. 3A (*15*). The similar FC and ZFC $4\pi\chi_{dc}^{ab}$ observed above ~ 30 K in the samples for the 49 K transition as evident from Fig. 1C suggest that superconductivity should exist inside the platelets with mesoscopic structures described in the model and below, consistent with the proposition of interface superconductivity.

Recently, through a systematic compositional and magnetic study of $(Ca_{1-x}Pr_x)122$ single crystals (*18*), we observed the unexpected superparamagnetism through Pr-doping in the region of $0.6 < x < 0.13$, where the high $T_c$-superconductivity exists. The superparamagnetic cluster density $n$ associated with the superparamagnetism is found to scale well with the compositional defect density of As-vacancies. What is even more amazing is the similarity of x-dependences of the superconducting volume fraction *f* and *n*, shown in Fig. 2B. The superparamagnetism detected also shows a large anisotropy of ~ 7.5 at 5 K and 5 T, perpendicular vs. parallel to the c-axis, suggesting that the magnetic clusters cannot be a result of random magnetic impurities. As the x increases, the inter-cluster interaction evolves from antiferromagnetic through Curie to ferromagnetic. There clearly exists an intricate correlation among the nonbulk superconductivity with high $T_c$, the superparamagnetism, and ordered defects in the *ab*-planes of $(Ca_{1-x}Pr_x)122$. Although superconductivity, superparamagnetism, and defects in these crystals are induced by Pr-doping over a narrow region, the x-independent $T_c$ observed in this study and the suppression of *f* via high temperature annealing without changing x reported (*18*) suggested that the nonbulk-superconductivity in $(Ca_{1-x}Pr_x)122$ cannot be a direct result of variation of charge-carrier-density from chemical Pr-doping in the conventional sense. Instead, an alternative of interfacial mechanism is possible. The ordered defects of As-vacancies in the FeAs-planes detected in $(Ca_{1-x}Pr_x)122$ may provide a reasonable vehicle to form the proposed mesoscopic structures of interfaces to facilitate the realization of interfacial mechanism for the enhanced $T_c$ observed.

In conclusion, the present detailed chemical, structural, and magnetic studies on well-characterized single crystals have provided convincing evidence for highly anisotropic superconducting platelets and, combined with the detection of anisotropic regions of As defects, interface-enhanced $T_c$. This may also help account for the nonbulk superconductivity previously reported in the undoped $AFe_2As_2$ (*3-5*) and one unit-cell epitaxial FeSe-films (*19*). The full deployment of the interface mechanism may lead to higher $T_c$ in other artificially made or naturally assembled material systems. A microscopic investigation of these defects will be needed to reveal the working of the interfaces, which would also be a notable achievement by itself as noted previously (*20*).

**Acknowledgments:** The helpful discussion with J. E. Hoffman of Harvard University during the later part of this work is greatly appreciated. The work in Houston is supported in part by U.S. Air Force Office of Scientific Research Grant No. FA9550-09-1-0656, U.S. Air Force Research Laboratory subcontract R15901 (CONTACT) through Rice University, the T. L. L. Temple Foundation, the John J. and Rebecca Moores Endowment, and the State of Texas through the Texas Center for Superconductivity at the University of Houston.

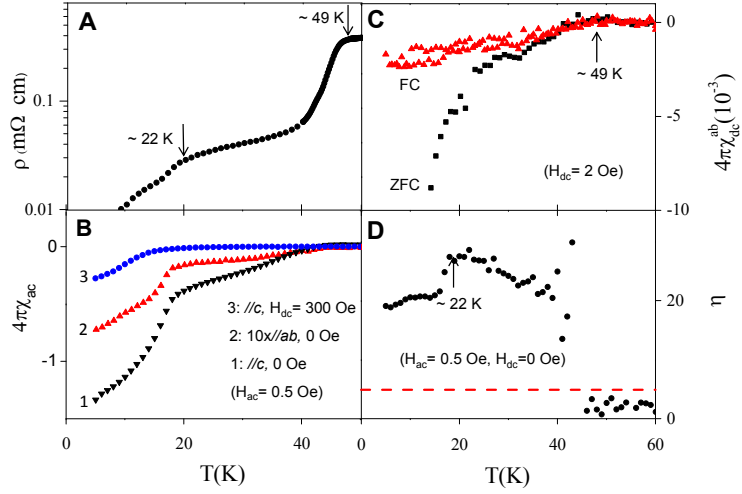

**Fig. 1**. Temperature dependences for the nominal $Ca_{0.87}Pr_{0.13}Fe_2As_2$ single crystal #1 of: **(A)** resistivity; **(B)** ac susceptibilities parallel to *ab*- and *c*-directions; **(C)** dc susceptibility; and **(D)** magnetic anisotropy $\eta(T) = \chi_{ac}^{c}/\chi_{ac}^{ab}$. The red dashed line represents the sample geometric anisotropy.

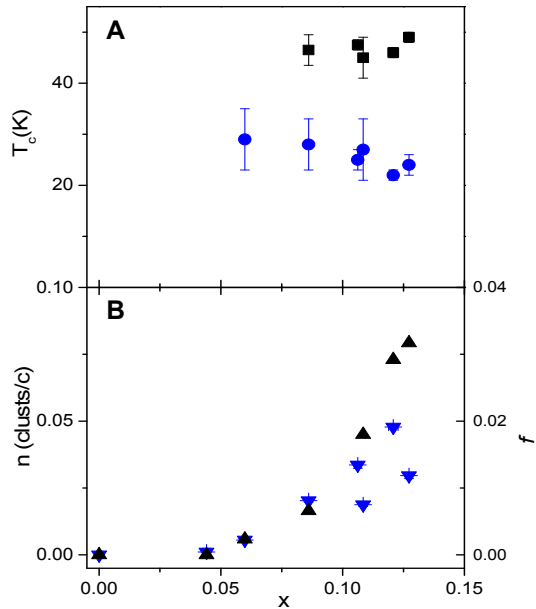

**Fig. 2.** The doping (x) dependences for nominal $(Ca_{1-x}Pr_x)Fe_2As_2$ single crystals of: **(A)** $T_c$ of the high-temperature transition (black squares) and the low temperature transition (blue circles) determined resistively; and **(B)** *f*, the superconducting volume fraction determined magnetically at 5 K (black triangles) and n, the superparamagnetic magnetic clusters per cell (blue inverted triangles).

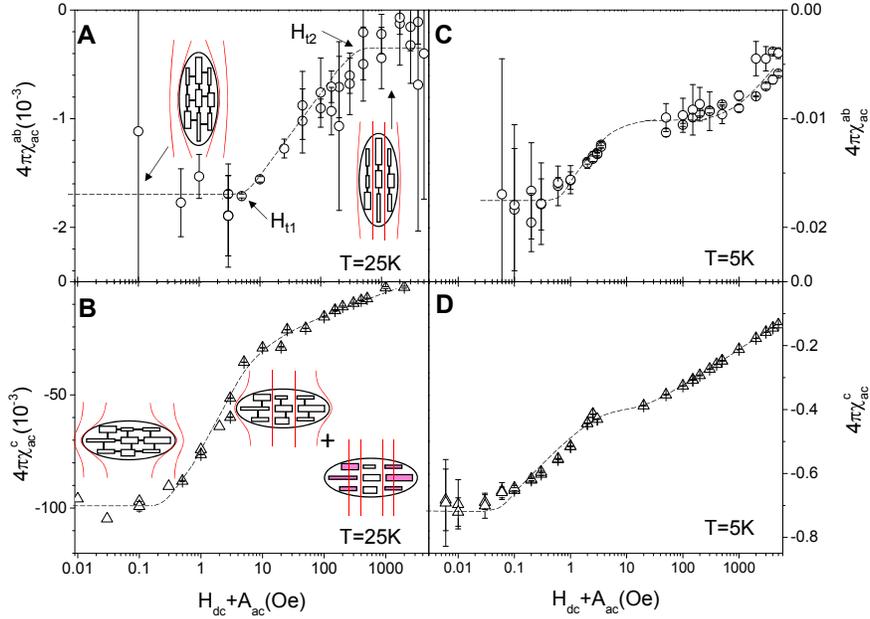

**Fig. 3.** The field effects on the ac magnetic susceptibilities of the nominal $Ca_{0.87}Pr_{0.13}Fe_2As_2$ single crystal #2 along the ab- and c-directions in a total fields $H_t = H_{ac} + H_{dc}$ between $5 \times 10^{-2}$ and $5 \times 10^3$ Oe at 25 K (**A**,**B**) and at 5 K (**C**,**D**). Red "bricks" represent the superconducting platelets after penetration by Abrikosov vortices.

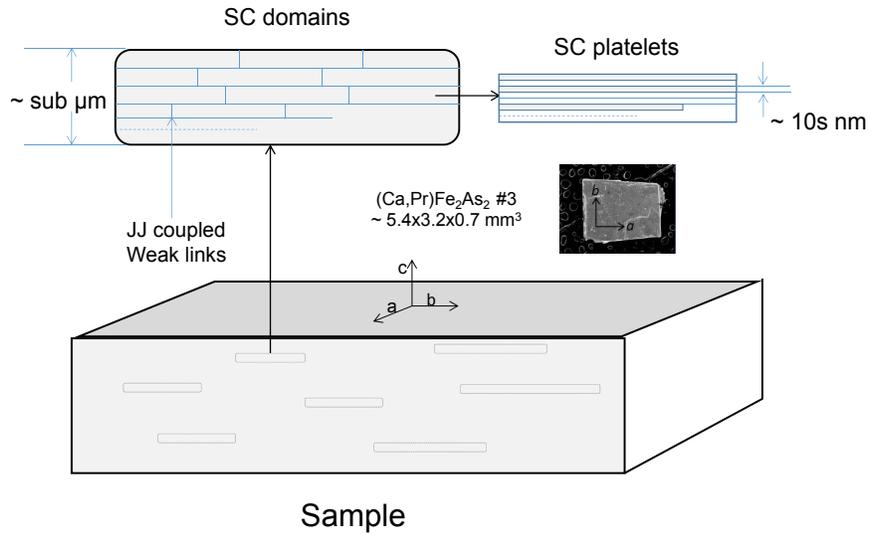

**Fig. 4.** The proposed model of naturally occurring superconducting morphology suggested by the unusual field penetration (Fig. 3) in the (Ca,R)122 single crystals with unusually high $T_c^{on}$. It consists of sparsely populated domains comprising Josephson-coupled superconducting platelets with large magnetic anisotropy. Within the platelets, different $Fe_2As_2$/defect-infested $Fe_2As_2$ planes may alternately stack and create the interfaces for enhanced $T_c$. The image of one sample investigated is also included.

**Supplementary Materials:**

Supplementary Text

Figs. S1-S3

References (*21-30*)

**Supplementary Materials:**

**S1. The 49 K -$T_c^{on}$ of (Ca,RE)Fe$_2$As$_2$ is higher than those of the doped and undoped AFe$_2$As$_2$ family with A = Ca, Sr, and Ba at ambient and under pressure; and any known compounds that contain one or more elements of Ca, RE, Fe, and As:**

    **(Ca,RE)Fe$_2$As$_2$**

        (Ca,La)Fe$_2$As$_2$          ≤ 43 K (non bulk) (present work)

        (Ca,Ce)Fe$_2$As$_2$          ≤ 42 K (nonbulk) (present work)

        (Ca,Pr)Fe$_2$As$_2$          ≤ 49 K (nonbulk) (present work)

        (Ca,Nd)Fe$_2$As$_2$          ≤ 42 K (nonbulk) (present work)

    **CaFe$_2$As$_2$**

        ambient                ~ 10 K (trace) (*17*)

        high pressure        ~ 12 K (filamentary) (*18,19*), ~ 8 K (bulk) (*20*)

        doping (with Na)    ~ 33 K (bulk) (*21*)

    **SrFe$_2$As$_2$**

        ambient                ~ 22 K (nonbulk) (*22*)

        high pressure        ~ 37.6 K (bulk?) (*23-25*)

        doping (with K)     ~ 31 K (bulk) (*26*)

    **BaFe$_2$As$_2$**

        ambient                ~ 23 K (nonbulk) (*27*)

        pressure              ~ 35 K (bulk?) (*23,28*)

        doping (with K)     ~ 38 K (bulk) (*29*)

    **Other known compounds that comprise two or more of the Ca, Pr, Fe, As elements in the (Ca,Pr)Fe$_2$As$_2$ superconductor are:**
    PrFe$_4$As$_{12}$, Pr$_2$Fe$_{17}$, PrFe$_2$, PrFe$_7$, PrAs, Pr$_4$As$_3$, CaAs, CaAs$_3$, Ca$_2$As$_3$, Ca$_2$As, Ca$_5$As$_3$, FeAs, FeAs$_2$, Fe$_2$As, and Fe$_{12}$As$_5$, none of which has been reported to be superconducting (*30*).

**S2. The unusually high $T_c$ has nothing to do with the collapsed phase transition.** All (Ca$_{1-x}$R$_x$)122 crystals with R = La, Ce, Pr, and Nd exhibit two superconducting transitions with $T_{c1}$ = 49-47 K and $T_{c2}$ = 24-14 K, whether they undergo a collapsed phase transition or not.

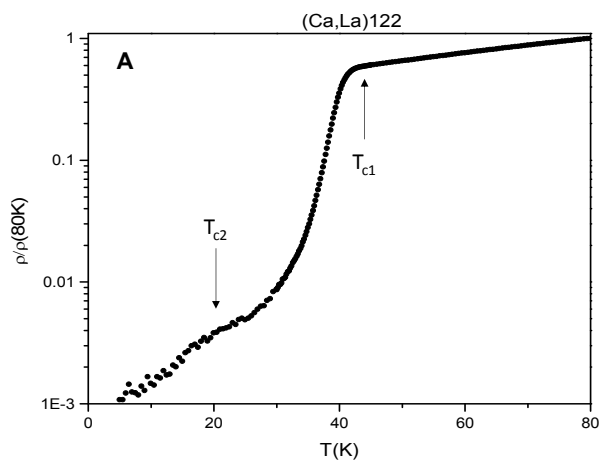
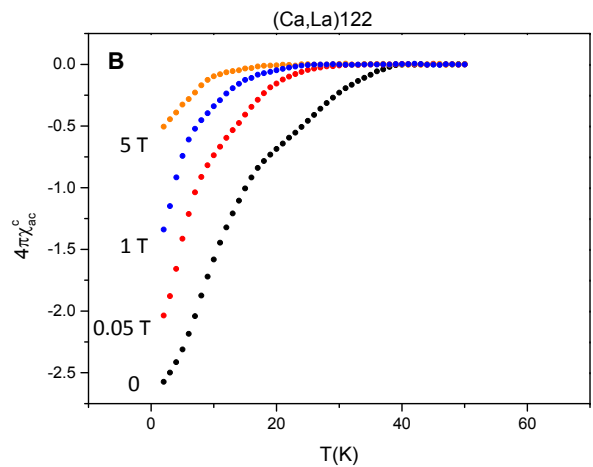
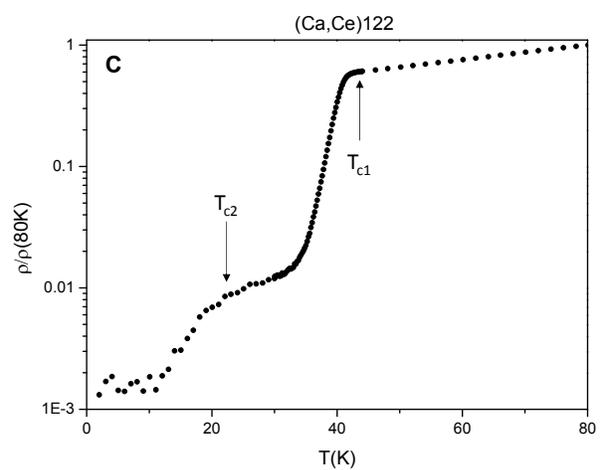
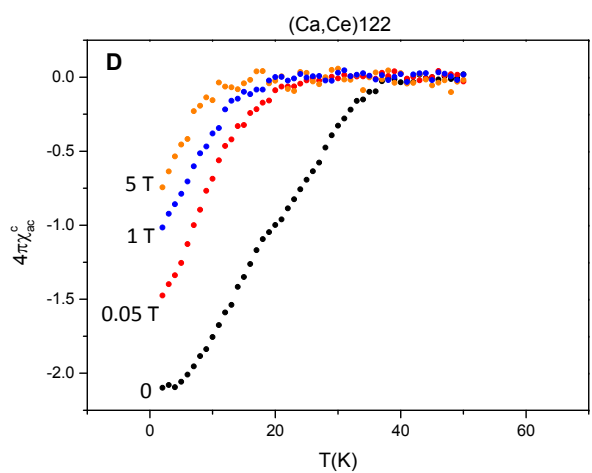
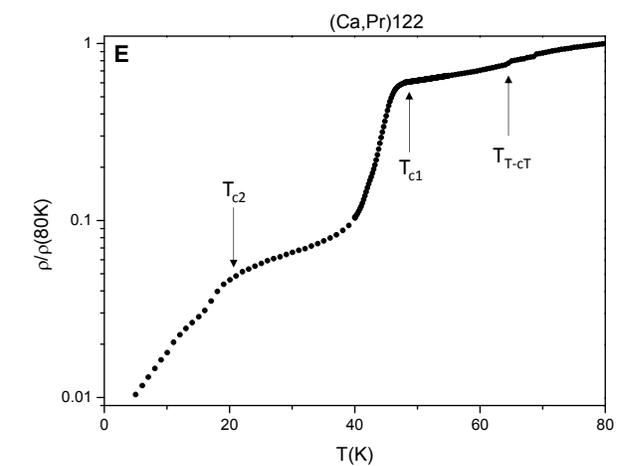
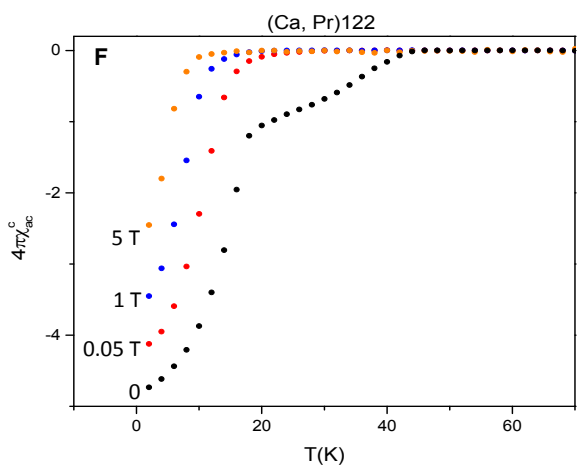

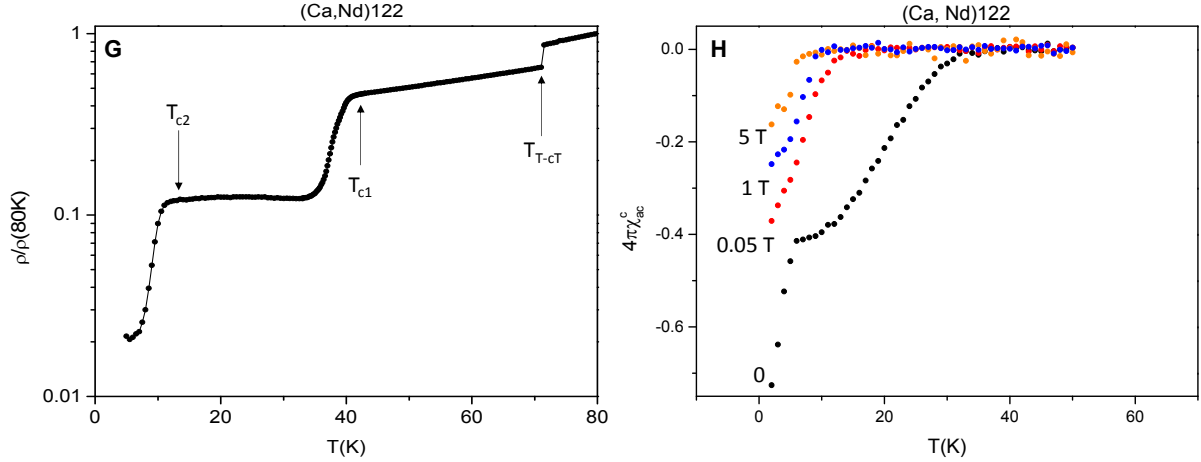

**Fig. S1.** The temperature dependences of risistivity and the ac magnetic susceptibility along the c-axis for (Ca,R)122 crystals with arrows showing the $T_c$ s. Note the suppression of the $T_{c1}$-transition by a field of ~ 500 Oe, in contrast to the $T_{c2}$-transition that survies 5 T.

**S3. The two superconducting transitions in $(Ca_{1-x}Pr_x)122$ crystals evolve discontinuously with x, consistent with data in Fig. 2A.**

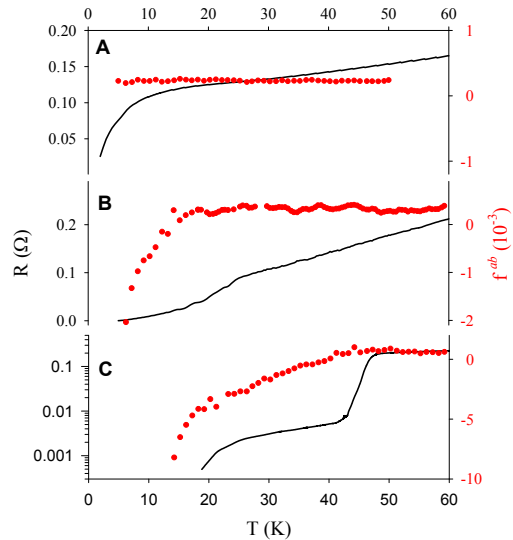

**Fig. S2.** Temperature-dependence of $f \equiv |4\pi \chi_{ac}^{ab}|$ (red) and R (black) for crystals with x = 0.04 (**A**), 0.06 (**B**), and 0.11 (**C**). The small R-drop around 10 K can be attributed to the small superconducting signal of the undoped Ca122.

**S4. Although WDS shows no Sm or Gd in crystals nominally doped with them, the crystals together with the growth flux were examined to determine if the nonbulk superconductivity with high $T_c$ is caused by small inclusion of $RFeAsO_{1-\delta}$. No superconductivity was found above 15 K, except trace of superconductivity below 15 K attributable to undoped Ca122 (*17*).**

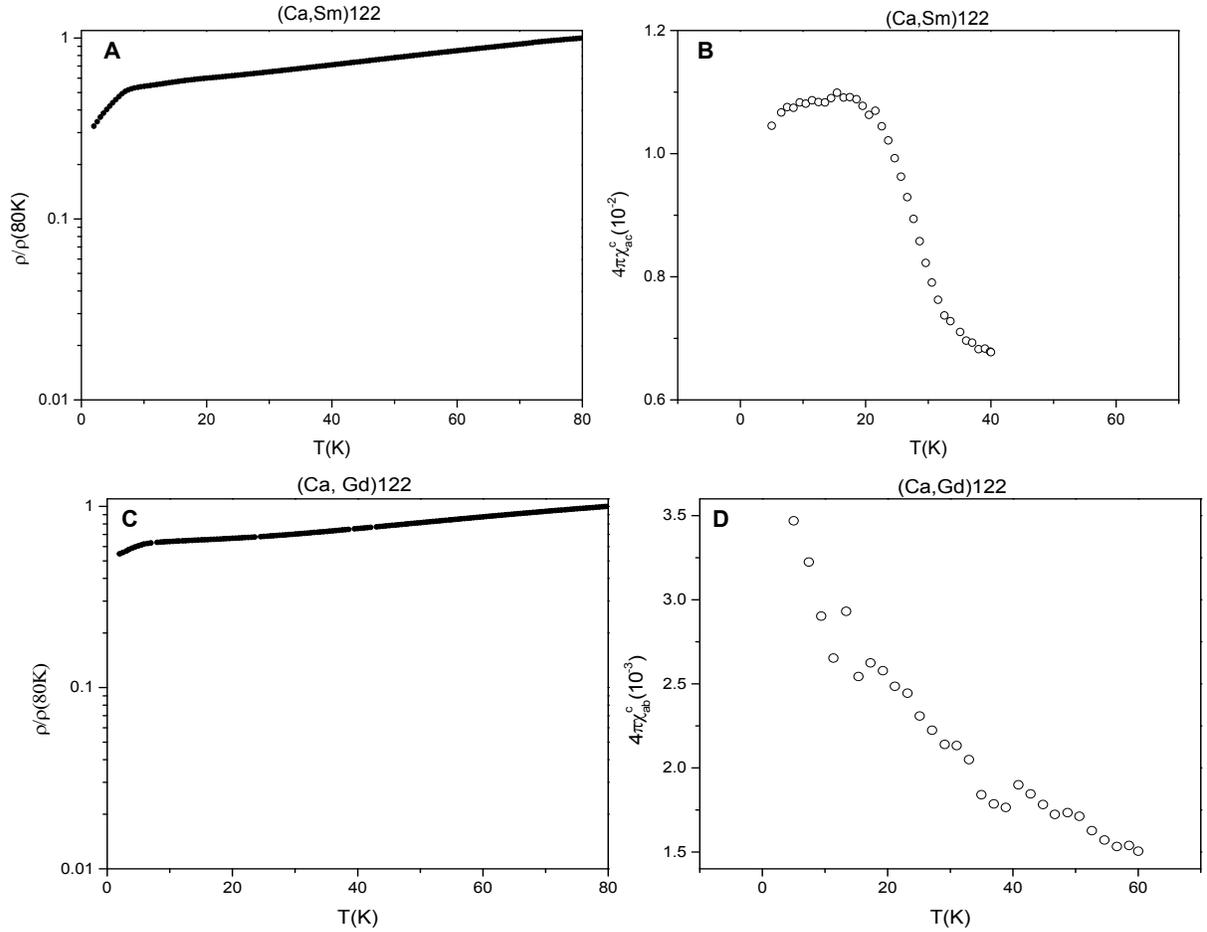

**Fig. S3.** Temperature dependences of resistivity and dc magnetic susceptibility of crystals (Ca,R)122 together with the crystal growth fluxes with R = Sm and Gd in an attempt not to miss any possible RFeAsO$_{1-\delta}$ that superconducts up to 56 K (*30*). No superconductivity was detected above 10 K except a very small trace in resistivity below 10 K attribuable to undoped Ca122 (*17*), ruling out possible RFeAsO$_{1-\delta}$ inclusion to be the culprit for the high T$_c$ nonbulk-superconductivity observed in other (Ca,R)122.